\documentclass[%
 reprint,
amsmath,amssymb,
aps,
prb,
]{revtex4-2}
\usepackage{multirow}
\usepackage{amsthm}
\usepackage{amsmath}
\usepackage{amsfonts}
\usepackage[ruled,vlined]{algorithm2e}
\usepackage{comment}
\usepackage{hyperref}
\usepackage{color}
\usepackage{tabularx}
\usepackage{graphicx}
\usepackage{bbm}
\usepackage{float}
\usepackage[normalem]{ulem}

\newtheorem{defn}{Definition}[section]

\newcommand{\dsep}{D_\mathrm{sep}}

\usepackage{graphicx}
\usepackage{dcolumn}
\usepackage{bm}

\newcommand{\ket}[1]{|#1\rangle}

\begin{document}

\author{Lexin Ding}
\affiliation{Faculty of Physics, Arnold Sommerfeld Centre for Theoretical Physics (ASC),\\Ludwig-Maximilians-Universit{\"a}t M{\"u}nchen, Theresienstr.~37, 80333 M{\"u}nchen, Germany}
\affiliation{Munich Center for Quantum Science and Technology (MCQST), Schellingstrasse 4, 80799 M{\"u}nchen, Germany}

\author{Gesa D{\"u}nnweber}
\affiliation{Faculty of Physics, Arnold Sommerfeld Centre for Theoretical Physics (ASC),\\Ludwig-Maximilians-Universit{\"a}t M{\"u}nchen, Theresienstr.~37, 80333 M{\"u}nchen, Germany}
\affiliation{Munich Center for Quantum Science and Technology (MCQST), Schellingstrasse 4, 80799 M{\"u}nchen, Germany}

\author{Christian Schilling}
\email{c.schilling@lmu.de}
\affiliation{Faculty of Physics, Arnold Sommerfeld Centre for Theoretical Physics (ASC),\\Ludwig-Maximilians-Universit{\"a}t M{\"u}nchen, Theresienstr.~37, 80333 M{\"u}nchen, Germany}
\affiliation{Munich Center for Quantum Science and Technology (MCQST), Schellingstrasse 4, 80799 M{\"u}nchen, Germany}

\title{Physical Entanglement Between Localized Orbitals}

\begin{abstract}
The goal of the present work is to guide the development of quantum technologies in the context of fermionic systems. For this, we first elucidate the process of entanglement swapping in electron systems such as atoms, molecules or solid bodies. This demonstrates the significance of the number-parity superselection rule and highlights the relevance of localized few-orbital subsystems for quantum information processing tasks. Then, we explore and quantify the entanglement between localized orbitals in two systems, a tight-binding model of non-interacting  electrons and the hydrogen ring. For this, we apply the first closed formula of a faithful entanglement measure, derived in [arXiv:2207.03377] as an extension of the von Neumann entropy to genuinely correlated many-orbital systems. For both systems, long-distance entanglement is found at low and high densities $\eta$, whereas for medium densities, $\eta \approx \frac{1}{2}$, practically only neighboring orbitals are entangled. The Coulomb interaction does not change the entanglement pattern qualitatively except for low and high densities where the entanglement increases as function of the distance between both orbitals.
\end{abstract}

\maketitle

\section{Introduction}







Entanglement is an essential resource for quantum information processing (QIP)\cite{wootters1998quantum} and it has been intensively studied in the context of \emph{distinguishable} particles. In the era of the second quantum revolution\cite{deutsch2022second,dowling2003quantum}, however, the new primary platforms for executing QIP tasks will likely comprise atoms and molecules. It is therefore of utmost importance to identify and quantify the \textit{physical entanglement} (that which can be physically accessed), for systems of \emph{indistinguishable} particles, particularly electrons.



That being said, quantifying entanglement in electronic systems is by no means a straightforward task. Even more urgently, there is a misalignment between the common treatment and quantification of electronic entanglement\cite{Reiher12,Reiher15,Gigena15,Reiher16,Pachos17free,aoto2020grassmannians,benatti2020entanglement,pusuluk2022classical,faba2021correlation,faba2021two} and the goal of utilizing it for QIP tasks\cite{franco2018indistinguishability,olofsson2020quantum,debarba20teleport,morris2020entanglement,galler2021orbital}. It is therefore the motivation of our present work to communicate and elucidate three key aspects which are preferable if not indispensable for comprehensively achieving the latter: (i) \textit{localized molecular orbitals}, on which local operations could be performed to realize QIP tasks, (ii) \textit{operationally meaningful entanglement quantification}, which evaluates precisely the useful entanglement for QIP, and (iii) \textit{computable measures for entanglement}.

To recognize the underdevelopment of point (i), we recall that common orbitals used in quantum chemistry are typically not localized. Indeed, by forming delocalized molecular orbitals one can reduce the complexity of the electronic ground state problem since these orbitals tend to correlate less with one another compared to fully-localized atomic orbitals. In fact,
optimizing the orbital basis to lower the correlation and thus the computational cost in solving the ground state problem has become a hot topic\cite{Legeza03,Chan08,Chan12,Legeza15Rev}. However, this objective is in direct contradiction with our ambition to use orbital entanglement for QIP purposes. A large family of QIP protocols uses entanglement as a resource to overcome spatial separation\cite{ekert1991quantum,bennett1992communication,bennett1992experimental,mattle1996dense,bouwmeester1997experimental,bennett2020quantum}, for instance, teleporting a quantum state. If we are to realize these protocols on a molecule, we must head in the direction opposite to the common approach of delocalizing molecular orbitals. The entanglement needs to be as high as possible and distributed across separated \textit{local} regions, allowing two \emph{local} agents to harness it with \emph{local} operations. To achieve this, one must construct and utilize orbitals localized to as few atomic centers as possible, which inevitably comes with a significant computational cost.

The necessity of point (ii) originates from the superselection rules\cite{SSR,wick1970superselection} (SSR) for fermions. Performing QIP tasks with electronic orbitals essentially comes down to implementing local operations on individual orbitals\cite{krylov20oribitals}. Yet, not all local operations can be realized due to nature's restriction ensuring that the number-parity is locally preserved\cite{SSR}. This in turns implies that entanglement arising from breaking the number parity, the so-called ``fluffy bunny'' entanglement\cite{wise04fluffy}, cannot be utilized for our purposes. Ignoring in concrete calculations the SSR would lead to a gross overestimation of the physical entanglement\cite{bartlett2003entanglement,banuls07fermion,ding2020correlation}, the entanglement that \textit{can} be utilized, and thus the working capacity of the electronic system at hand for QIP applications. Therefore the importance of addressing the effect of SSR on the available entanglement in the system can hardly be overemphasized.

As for (iii), although there is no shortage of meaningful definitions of entanglement measures, thanks to the abstract mathematical foundation provided by quantum information theory, they are mostly notoriously difficult to calculate. As a consequence, analytic treatment of orbital entanglement in many-body physics is almost always restricted to the so-called block entropy\cite{calabrese2009entropy,calabrese2011entropy,vidmar2017entropy}.
This quantity, despite its analytic computability\cite{peschel2003reduced}, suffers from two critical shortcomings. First, it quantifies the entanglement between the block and its complement \textit{only} when the overall system is in a pure state. Otherwise, it effectively measures the entanglement of the block and the rest of the universe. Second, it is invariant under unitary transformations within the block. In other words, it alone cannot capture any internal structure such as the entanglement between orbitals, which is the sought-after resource for QIP. To fill this vital gap, Ref.~\cite{ding2022quantifying} recently provided the first closed formula for the faithful entanglement between electronic orbitals.

It is therefore one of the key achievements of this paper to lay out and accentuate these three essential points (i)-(iii) to the quantum science and technology community. By referring to an entanglement swapping protocol, we demonstrate that the useful orbital entanglement for QIP tasks is quantified with SSR taken into account. To showcase the analytic tools provided in Ref.~\cite{ding2022quantifying}, we then study the exactly solvable tight-binding model. There, we break the usual paradigm of restricting analytic studies to the block entropy by performing a fully analytic analysis of the physical entanglement between any two spatial orbitals. Lastly, we upgrade the tight-binding model to the hydrogen ring and analyze the orbital-orbital entanglement therein. By tackling the numerical complexity from both the orbital localization and the Coulomb interaction, we manage to realize potent entanglement between two far separated fully-localized orbitals.

The paper is structured as follows: In Section \ref{sec:concepts}, we first review the relevant notions of fermionic entanglement. After establishing the mathematical framework, we demonstrate the necessity of SSRs by discussing an entanglement swapping protocol. Moreover, we restate the analytic formula for orbital entanglement. In Section \ref{sec:freefermions} we calculate the orbital-orbital entanglement in the tight-binding model and evaluate by analytical means its asymptotic behavior. Finally, in Section \ref{sec:hydrogen} we explore the entanglement pattern between localized orbitals in the ground state of the hydrogen ring.

\section{Concepts}

\label{sec:concepts}

In this Section, we first review the relevant quantum information concepts and explain how to operationally transfer these concepts to fermionic systems. Secondly, we demonstrate the effect of superselection rules on the accessible entanglement in fermionic systems, by invoking an entanglement swapping protocol. Lastly, we briefly restate the entanglement formula for two orbitals derived in Ref.~\cite{ding2022quantifying}.

\subsection{Locality and entanglement}

The notion of locality is the starting point for the theory of entanglement. We consider two distant parties, Alice and Bob, each possessing a local quantum system with a finite-dimensional Hilbert space denoted by $\mathcal{H}_A$ and $\mathcal{H}_B$, respectively. The two systems are local in the sense that an action of one party on their subsystem has no effect on the other. Mathematically speaking, this separation is described by the tensor product structure of both the total Hilbert space $\mathcal{H} = \mathcal{H}_A \otimes \mathcal{H}_B$ and the total algebra of observables $\mathcal{A} = \mathcal{A}_A \otimes \mathcal{A}_B$.

With the stage set, a quantum state $\rho_{AB}$ shared by Alice and Bob is said to be separable, if and only if it can be prepared by local operations and classical communication (LOCC)\cite{werner89}. Local operations allow Alice and Bob to prepare any product state $\rho_A \otimes \rho_B$, and classical communication can be used to generate convex mixtures of those states. This then gives rise to the convex set of bipartite separable states $D_\mathrm{sep}$:
\begin{defn}[Bipartite Separability\cite{werner89}]
A bipartite state $\rho_{AB}$ on $\mathcal{H}_A\otimes\mathcal{H}_B$ is separable if and only if it can be expressed as
\begin{equation}
    \rho_{AB} = \sum_i \,p_i\, \rho_A^{(i)} \otimes \rho_B^{(i)}
\end{equation}
with $p_i \geq 0$ and $\sum_i p_i = 1$. The set of separable states is denoted as $\dsep$. Any state that is not in $\dsep$ is necessarily entangled.
\end{defn}
A non-separable, or entangled state contains valuable entanglement resource for QIP tasks\cite{ekert1991quantum,bennett1992communication,bennett1992experimental,mattle1996dense,bouwmeester1997experimental,bennett2020quantum}. This resourcefulness of entanglement is distinct from the mutual information, which contains both quantum and classical correlation\cite{modi2010unified}. Therefore the detection and quantification of entanglement in a bipartite system is of crucial interest.
For a pure state $\rho_{AB} = |\Psi_{AB}\rangle\langle\Psi_{AB}|$, it is easy to verify whether or not it belongs to $\dsep$. It suffices to check if it is a pure product state. Or equivalently one can check the purity of the reduced state $\rho_{A/B} = \mathrm{Tr}_{B/A}[|\Psi_{AB}\rangle\langle\Psi_{AB}|]$. The degree of mixedness of $\rho_{A/B}$ thus accounts for the entanglement in $\rho_{AB}$. This leads to a well-defined entanglement measure for \textit{pure states}
\begin{equation}
    E(|\Psi_{AB}\rangle\langle\Psi_{AB}|) = S(\rho_{A/B}) \label{eqn:pure_ent},
\end{equation}
where $S(\rho) = -\mathrm{Tr}[\rho\log(\rho)]$ is the von Neumann entropy. For mixed states, however, there is no such straightforward analytic check for separability, except for qubit-qubit and qubit-qutrit systems\cite{peres1996separability,horodecki1996separability}. In fact, even the problem of determining the separability of a general mixed quantum state is NP-hard\cite{gurvits03np,gharibian2008np}, and in practice can only be approximately solved with significant computational efforts using symmetric extension and semidefinite programming\cite{doherty04separability,plenio09separability}. On the formal level, the degree of entanglement of a mixed state can be defined geometrically as the distance to the closest separable state (here and in the following we occasionally skip the index $AB$ of the joint state)
\begin{equation}
    E(\rho) = \min_{\sigma \in \dsep} S(\rho\|\sigma) \label{eqn:rel_ent},
\end{equation}
where the generalized distance function is chosen to be the relative entropy
\begin{equation}
    S(\rho\|\sigma) = \mathrm{Tr}[\rho(\log(\rho)-\log(\sigma))]
\end{equation}
for its rich quantum informational meaning. To be more precise, the quantum relative entropy $S(\rho\|\sigma)$ exponentially suppresses the ``difficulty'' of telling $\rho$ and $\sigma$ apart\cite{vedral98purification}, in the spirit of the so-called quantum Sanov's theorem\cite{vedral98purification}. The measure \eqref{eqn:rel_ent} is called the relative entropy of entanglement\cite{henderson2000information}, and will be used as the standard entanglement measure throughout this paper. When applied to pure states, \eqref{eqn:rel_ent} reduces to the pure state measure via the von Neumann entropy \eqref{eqn:pure_ent}\cite{vedral98purification}.

\subsection{Fermionic locality and superselection rules}

In the previous section, we explained that the notion of entanglement necessitates two ingredients of locality: (1) a tensor product structure of the Hilbert space, and (2) a tensor product structure of the algebra of observables. In the context of fermionic systems, this raises the following questions. Does there also exist a tensor product structure in a fully-antisymmetrized $N$-fermion Hilbert space, e.g. for a hydrogen molecule with two electrons? And how can one identify two commuting algebras of observables when fermionic creation/annihilation operators anticommute?

The fermionic analogue of the bipartition into distant laboratories corresponds to a choice of splitting the one particle Hilbert space $\mathcal{H}_1$ (or a subspace of it) into two orthogonal subspaces. For instance, $\mathcal{H}_1$ can be realized by an array of Wannier orbitals, or an arrangement of molecular spin orbitals. From here on we shall refer to the elements of an orthonormal basis of $\mathcal{H}_1$ as orbitals. A partition of an orbital basis $\{|\varphi_i\rangle\}_{i=1}^{d}$ into two complementing subsets $A$ and $B$ dissolves the one particle Hilbert space into a direct sum $\mathcal{H}_1 = \mathcal{H}_1^{(A)} \oplus \mathcal{H}_1^{(B)}$. This for example can correspond to separating a lattice into the left and right parts at some cut. Each part generates a local Fock space and together they form a tensor product of the total Fock space $\mathcal{F}(\mathcal{H}_1) \cong \mathcal{F}(\mathcal{H}_1^{(A)}) \otimes \mathcal{F}(\mathcal{H}_1^{(B)})$ through the following basis identification
\begin{equation}
    |n_1,n_2,\ldots,n_d\rangle = |n_1, \ldots, n_{d_A} \rangle \otimes |n_{d_A+1},\ldots ,n_d\rangle,
\end{equation}
where $n_1,\ldots,n_d \in \{0,1\}$,
\begin{equation}\label{eq:conf}
\begin{split}
    |n_1,\ldots,n_d\rangle \equiv (f^\dagger_{1})^{n_1}\cdots(f^\dagger_{d})^{n_d}|0\rangle,
    \end{split}
\end{equation}
and $f^{(\dagger)}_{i}$ denotes the operator that annihilates (creates) a fermion in state $|\varphi_i\rangle$.

As for the second necessary ingredient, i.e., commuting local algebras of observables, it seems at first sight to be in direct contradiction with the nature of fermionic operators. Suppose Alice and Bob each possess a fermionic mode at the opposite ends of the universe. Due to the anticommutation relation between the creation operators, Alice's action could in principle still influence Bob's mode, and accordingly realize superluminal signaling\cite{johansson2016comment,ding2020concept}. This manifest violation of special relativity is in reality not possible, due to nature's restriction on the local operations by Alice and Bob, which can be formulated as so-called superselection rules\cite{wick1970superselection,SSR} (SSR). The relevant superselection rules depend on the conserved observables in the theory or even practical limitations. Nonetheless, the parity superselection rule\cite{SSR} (P-SSR) is ubiquitous: one cannot superpose quantum states with different local fermionic parity. Equivalently, physical local observables must always commute with the local parity operator\cite{bartlett2003entanglement,wise04fluffy,schuch04super,jones06super} $\hat{\Pi} \equiv \exp(i\pi\sum_l \hat{n}_l) = \hat{P}_+ - \hat{P}_-$ where the sum is over all local orbital degrees of freedom, $\hat{P}_\pm$ is the orthogonal projection onto the even($+$)/odd($-$) subspace, and $\hat{n}_l \equiv f_l^\dagger f_l$ denotes the particle number operator for mode $\ket{\varphi_l}$. The local observables $\hat{O}$ fulfilling this requirement are polynomials of $f^\dagger_i f^\dagger_j$, $f^{\phantom{\dagger}}_i f^{\phantom{\dagger}}_j$, and $f^{{\dagger}}_if^{\phantom{\dagger}}_j$, and satisfy
\begin{equation}
    \hat{O} = \mathcal{G}_{\Pi}[\hat{O}]\equiv \hat{P}_+ \hat{O} \hat{P}_+ + \hat{P}_- \hat{O} \hat{P}_- \label{eqn:cov_O},
\end{equation}
where $\mathcal{G}_\Pi$ denotes the projective map onto the P-SSR compatible sub-algebra. Then, it follows directly that any two observables $\hat{A}$ and $\hat{B}$ referring to two distinct sets of fermionic modes and satisfying locally \eqref{eqn:cov_O} necessarily commute with each other. In other words, algebras of parity preserving observables on two distinct fermionic subsystems do exhibit a tensor product structure. More generally, any physical quantum channel $\mathcal{E}$ (completely positive and trace-preserving map\cite{nielsen2002quantum}) on the reduced state must satisfy
\begin{equation}
    \mathcal{E} = \mathcal{G}_\Pi \circ \mathcal{E} \circ \mathcal{G}_\Pi.
\end{equation}

\subsection{Entanglement swapping under SSR}

Nature's restriction on the implementable local observables has a profound implication on the entanglement in fermionic states. From the perspective of state tomography, coherent terms between different local parity states can never be observed under P-SSR, for the observable needed for such measurement is simply not available by the laws of physics. Therefore, the entanglement arising from SSR-violating coherence, also known as the ``fluffy bunny entanglement''\cite{wise04fluffy}, is not accessible. This reduction of entanglement due to SSR is proven rigorously in Ref.~\cite{bartlett2003entanglement} on an abstract mathematical level. Here we demonstrate it in concrete terms with a swapping protocol.

\begin{figure}[t]
    \centering
    \includegraphics[scale=0.35]{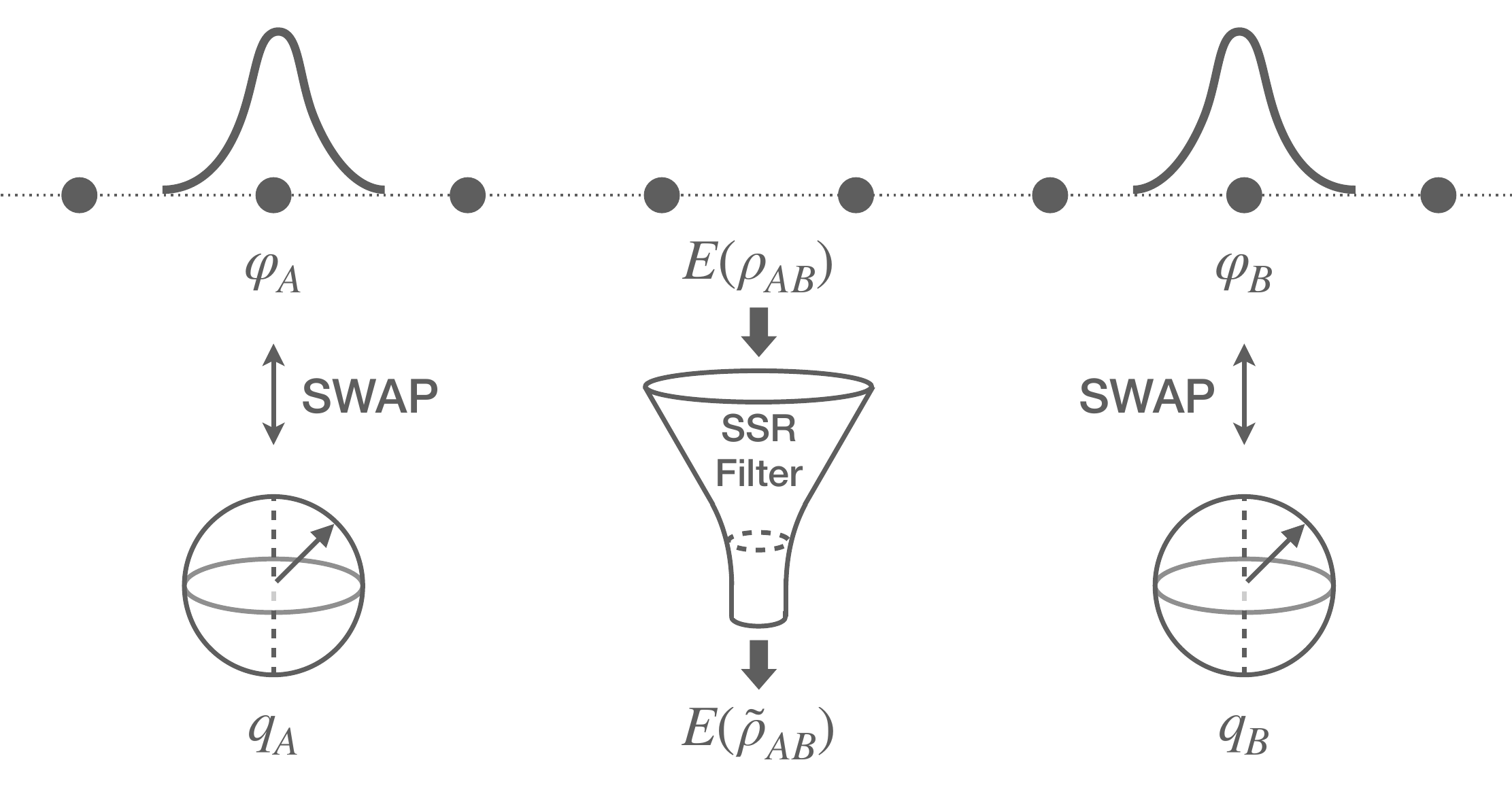}
    \caption{Entanglement swapping protocol under SSR between the two electronic orbitals $\varphi_A$ and $\varphi_B$, and the two qubit registers $q_A$ and $q_B$. See text for more details.}
    \label{fig:ent_swapping}
\end{figure}

In Figure \ref{fig:ent_swapping} we depict schematically an entanglement swapping protocol between two fermionic orbitals $(\varphi_A,\varphi_B)$ in the state $\rho_{AB}$, and two qubits $(q_A,q_B)$ in the state $|00\rangle$, both shared by Alice and Bob. We shall denote the orbital Fock spaces as $\mathcal{F}_{\varphi_{A/B}}$ and the qubit Hilbert spaces as $\mathcal{H}_{q_{A/B}}$. For both the Fock spaces and the qubit Hilbert spaces, we denote the local basis as $\mathcal{B} = \{|0\rangle,|1\rangle\}$. One should keep in mind that for the qubit space $\mathcal{B}$ is the computational basis, while for the Fock space, $\mathcal{B}$ refers to the occupational basis. The protocol in Figure \ref{fig:ent_swapping} extracts the entanglement between the fermionic modes onto the qubit registers, on which this entanglement could be used to perform more flexible quantum information protocols.

The protocol consists of Alice swapping the two states on $\varphi_A$ and $q_A$, and Bob doing the same on $\varphi_B$ and $q_B$. Specifically, the SWAP channel $\mathcal{S}$ on a bipartite system is defined element-wise as
\begin{equation}
    \mathcal{S}\!: \,\,|i\rangle \langle j | \otimes |k\rangle\langle l | \mapsto |k\rangle \langle l | \otimes |i\rangle\langle j |,
\end{equation}
for all $i,j,k,l\in\{0,1\}$.
In a world without SSR, after the SWAP operation, the final state of the two qubits would be exactly $\rho_{AB}$ and it would accordingly contain exactly the same amount of entanglement as the original two-orbital state. However, the SWAP operation clearly does not preserve the local parities, and hence cannot be realized with physical fermionic operators. Instead, Alice/Bob can only perform the superselected channel
\begin{equation}
    \tilde{\mathcal{S}}_{A/B} = \mathcal{G}_{\Pi_{A/B}} \circ \mathcal{S}_{A/B} \circ \mathcal{G}_{\Pi_{A/B}}. \label{eqn:swap_ssr}
\end{equation}

To see the difference between $\mathcal{S}_{A/B}$ and $\tilde{\mathcal{S}}_{A/B}$, let us first observe the action of the SWAP channel on a pure state $|\psi_A\rangle = |+\rangle\otimes|+\rangle \in \mathcal{F}_{\varphi_A} \otimes \mathcal{H}_{q_A}$ where $|+\rangle = \frac{1}{\sqrt{2}}(|0\rangle+|1\rangle)$, which clearly violates the P-SSR. In a world free of SSR, $|\psi_A\rangle$ would be invariant under the SWAP channel. Under the P-SSR, however, the action of the physically implementable $\Tilde{\mathcal{S}}_A$ consists of three steps: (1) $\mathcal{G}_{\Pi_A}$ turns $|\psi_A\rangle$ into a mixed state $ \mathcal{G}_{\Pi_A} [\rho_A] = \frac{\openone}{2} \otimes |+\rangle\langle+|$. Namely, the coherent terms between different fermionic parities are erased. (2) $\mathcal{S}_A$ swaps the two tensor states to $\mathcal{S}_A \circ \mathcal{G}_{\Pi_A} [\rho_A] = |+\rangle\langle+| \otimes \frac{\openone}{2}$, which again contains P-SSR violating terms. (3) $\mathcal{G}_{\Pi_A}$ then eliminates said terms and the final state is $\mathcal{G}_{\Pi_A} \circ \mathcal{S}_A \circ \mathcal{G}_{\Pi_A} [\rho_A] = \frac{\openone}{2}\otimes\frac{\openone}{2}$. In this example we see evidently that SWAP does not leave $|\psi_A\rangle$ invariant, despite its apparent symmetry. Instead, P-SSR filters out any coherence between different parity sectors, and only the superselected state $\tilde{\rho}^\mathrm{P}_A \equiv \mathcal{G}_{\Pi_A}[|\psi_A\rangle\langle\psi_A|]$ can be transferred onto the qubit register. Moreover, the final state on the fermionic mode $\varphi_A$ is also the superselected variant of the original qubit state.

Now that we understand the action of \eqref{eqn:swap_ssr}, we can proceed to write down the action of the protocol $\tilde{S} \equiv \tilde{\mathcal{S}}_A \otimes \tilde{\mathcal{S}}_B$ on the composite Hilbert space $\mathcal{F}_{\varphi_A} \otimes \mathcal{H}_{q_A} \otimes \mathcal{F}_{\varphi_B} \otimes \mathcal{H}_{q_B}$: for any states $\rho_{AB} $ on $ \mathcal{F}_{\varphi_{A}} \otimes \mathcal{F}_{\varphi_{B}}$ and $\sigma_{AB} $ on $ \mathcal{H}_{q_{A}} \otimes \mathcal{H}_{q_{B}}$, we have
\begin{equation}
    \begin{split}
        \tilde{\mathcal{S}}[\rho_{AB} \otimes \sigma_{AB}] = \tilde{\sigma}^{\mathrm{P}}_{AB} \otimes \tilde{\rho}^{\mathrm{P}}_{AB}
    \end{split},
\end{equation}
where
\begin{equation}
    \begin{split}
        \tilde{\rho}^\mathrm{P}_{AB} &\equiv \mathcal{G}_{\Pi_A} \otimes \mathcal{G}_{\Pi_B}[\rho_{AB}]
        \\
        &= \sum_{s,s' = +,-} \hat{P}^{(A)}_s \otimes \hat{P}^{(B)}_{s'} \rho_{AB}\hat{P}^{(A)}_s \otimes \hat{P}^{(B)}_{s'}
    \end{split}
\end{equation}
and the same for $\tilde{\sigma}^\mathrm{P}_{AB}$. The final state $\tilde{\rho}^{\mathrm{P}}_{AB}$ on the qubit registers contains the same information as the original two-orbital state $\rho_{AB}$, except it is rid of the coherence between the different local parity sectors, which does not contribute to the physical entanglement. The decohering map $\mathcal{G}_{\Pi_A} \otimes \mathcal{G}_{\Pi_B}$ thus acts as an entanglement filter as depicted in Figure \ref{fig:ent_swapping}, and effectively reduces the available entanglement in the fermionic state. Only the entanglement within each local parity sector can be extracted for operational purposes.


\subsection{Analytic formula for orbital-orbital entanglement}

In the previous section, we concluded that the \textit{physical} entanglement in a bipartite fermionic state $\rho_{AB}$ in the presence of P-SSR, is precisely the entanglement in its superselected variant $\tilde{\rho}^{\mathrm{P}}_{AB} = \mathcal{G}_{\Pi_A} \otimes \mathcal{G}_{\Pi_B}[\rho_{AB}]$ quantified in the usual manner without P-SSR. In this Section, we will briefly summarize the analytic formula for physical entanglement between two spatial orbitals in electronic systems, derived in Ref.~\cite{ding2022quantifying}.

Consider two orbitals $A$ and $B$ with internal spin-$\frac{1}{2}$ degrees of freedom. This system is described by a state $\rho_{AB}$ acting on the Fock space $\mathcal{F}_A \otimes \mathcal{F}_B$, spanned by $\{|\alpha\rangle_A \otimes |\beta\rangle_B\}$, where $\alpha,\beta\in\{0,\uparrow,\downarrow,\uparrow\downarrow\}$. The mixedness of $\rho_{AB}$ accounts for the physically relevant scenario where $A$ and $B$ are in contact with other orbitals.
A key step in quantifying the entanglement $E(\tilde{\rho}^\mathrm{P}_{AB})$ is understanding its restriction to several smaller subspaces of the total Fock space, after imposing several common symmetries exhibited by \textit{realistic} condensed matter and quantum chemical systems.

The relevant systems in our case, the tight-binding model and the hydrogen ring, enjoy many symmetries, including (1) particle number, (2) magnetization and (3) reflection symmetry between orbitals $A$ and $B$ (e.g. two orthogonalized 1s orbitals in a hydrogen ring). Under these symmetries, there are only four entangled pure states compatible with local P-SSR:
\begin{equation}
    \begin{split}
        |\Phi_{\pm}\rangle &= \frac{|0\rangle_A \otimes |\!\uparrow\downarrow\rangle_B \pm |\!\uparrow\downarrow\rangle_A \otimes |0\rangle_B}{\sqrt{2}},
        \\
        |\Psi_{\pm}\rangle &= \frac{|\!\uparrow\rangle_A \otimes |\!\downarrow\rangle_B \pm |\!\downarrow\rangle_A \otimes |\!\uparrow\rangle_B}{\sqrt{2}}.
    \end{split}
\end{equation}
Naturally, if $\tilde{\rho}^{\mathrm{P}}_{AB}$ enjoys the symmetries (1), (2), and (3), then $|\Phi_\pm\rangle$ and $|\Psi_\pm\rangle$ are eigenstates of $\tilde{\rho}^{\mathrm{P}}_{AB}$, with eigenvalues $p_{\pm}$ and $q_{\pm}$, respectively. Even so, the exact P-SSR orbital-orbital entanglement, though entirely analytic, has a heavily involved form, which can be found in the Appendix of Ref.~\cite{ding2022quantifying}. Instead, we highlight another related entanglement formula from Ref.~\cite{ding2022quantifying} in the case where the particle number superselection rule (N-SSR) applies. In this setting, no superposition between different local particle number states is possible and therefore even $|\Phi_\pm\rangle$ are forbidden. The N-SSR compatible two-orbital state is obtained with the same reasoning via the superselection map
\begin{equation}
    \begin{split}
        \tilde{\rho}^\mathrm{N}_{AB} &\equiv \mathcal{G}_{N_A} \otimes \mathcal{G}_{N_B}[\rho_{AB}]
        \\
        &= \sum_{n_A,n_B=0}^{2} \hat{P}_{n_A,n_B} \rho_{AB} \hat{P}_{n_A,n_B},
    \end{split}
\end{equation}
where $\hat{P}_{n_A,n_B} = \hat{P}_{n_A} \otimes \hat{P}_{n_B}$ and $\hat{P}_{n_{A/B}}$ is the projection onto the sector of local particle number $n_{A/B}$ on subsystem $A/B$. The N-SSR entanglement of $\tilde{\rho}^\mathrm{N}_{AB}$ can then be derived, after considerable efforts, to be
\begin{equation}
    \begin{split}
        E(\tilde{\rho}^\mathrm{N}_{AB}) = \begin{cases} r\log\left(\frac{2r}{r+t}\right) + t\log\left(\frac{2t}{r+t}\right), \quad &r < t,
        \\
        0, \quad &r \geq t,
        \end{cases}\label{eqn:e_nssr}
    \end{split}
\end{equation}
where
\begin{equation}
    \begin{split}
        &t \equiv \max\{q_\pm\}, \quad r \equiv \mathrm{Tr}[\hat{P}_{1,1} \rho_{AB}] - t.
    \end{split}
\end{equation}
Note that $r < t$, or explicitly
\begin{equation}
    \mathrm{Tr}[\hat{P}_{1,1} \rho_{AB}] < 2\max\{q_\pm\} \label{eqn:ent_criteria},
\end{equation}
is the exact entanglement criterion. Below we will see that the P- and N-SSR orbital-orbital long-distance entanglement on a localized orbital chain are virtually indistinguishable. The simple form of the latter thus allows us to derive \textit{analytically} asymptotic properties of the physical entanglement between orbitals for both P- and N-SSR.

We remark that \eqref{eqn:e_nssr} is only applicable for states with the desired symmetry. Applying our formula to symmetry-broken states could lead to erroneous results. For example, according to \eqref{eqn:e_nssr}, the entanglement of $\frac{1}{\sqrt{2}}(|\!\!\uparrow\uparrow\rangle + |\!\!\downarrow\downarrow\rangle)$ would be zero, while this state is clearly entangled. However, one could apply a local spin flip (which preserves the entanglement), and arrive at the symmetric $|\Psi_+\rangle$ state, for which the entanglement formula \eqref{eqn:e_nssr} is applicable.

In the reminder of our paper we will apply the entanglement measure \eqref{eqn:e_nssr} and the corresponding one for the weaker P-SSR to two model systems in order to investigate their orbital-orbital entanglement as a function of, e.g., the electron density and the distance between both localized orbitals. This will provide new insights which may guide the utilization of the abundant entanglement resource for QIP tasks performed in realistic materials by tuning the particle density and interaction strength.

\section{Free fermions}\label{sec:freefermions}

The full benefit of having an analytic formula for orbital entanglement is best shown for systems where the ground state is also analytically available. Therefore, we turn to the exactly solvable free fermion models. Despite the absence of particle interaction, free fermion systems can exhibit various interesting phenomena such as topological phases\cite{ssh79,wen12topo,li14topo} and entanglement phase transition\cite{alberton21entanglement,zhang2022universal}. In this Section, we will focus on the tight-binding model and explore its entanglement pattern with our analytic tools. But instead of the commonly used block entropy\cite{calabrese2009entropy,calabrese2011entropy,vidmar2017entropy}, we will analytically derive the physical entanglement between any two spatial orbitals.

\subsection{Reduced states in free fermions systems}

A general particle number conserving non-interacting fermionic Hamiltonian can be written as a sum of one-body terms
\begin{equation}
    \hat{H}_\text{free} = \sum_{ij} h_{ij} f^\dagger_i f^{\phantom{\dagger}}_j, \quad h_{ij} = h_{ji}^\ast.\label{eqn:H_free}
\end{equation}
The spectrum of $H_\text{free}$ is analytically solved by finding a unitary matrix $U$ such that $\mathbf{d} = \mathbf{U} \mathbf{h} \mathbf{U}^\dagger$ is diagonal. We then rewrite $\hat{H}_\text{free}$ as
\begin{equation}
    \hat{H}_\text{free} = \sum_i d_{ii} c^{\dagger}_i c^{\phantom{\dagger}}_i \equiv \sum_i d_{ii} \hat{n}_i, \quad c^\dagger_i = \sum_j U_{ij} f^\dagger_j.
\end{equation}
The $N$-particle ground states of $\hat{H}_\text{free}$ is simply a Slater determinant state of the form
\begin{equation}
    |\Psi_N\rangle = c^\dagger_1c^\dagger_2\cdots c^\dagger_N|0\rangle
\end{equation}
with ground state energy $\sum_{i=1}^N d_{ii}$, assuming the $d_{ii}$'s are ordered increasingly.

To study the entanglement within the ground state $|\Psi_N\rangle$, a natural object to consider is the reduced density matrix $\rho_A$ of the ground state on a sub-lattice $A$. Its entropy quantifies the entanglement between the subsystem $A$ and its complement, and serves as a widely used analytic tool\cite{calabrese2009entropy,calabrese2011entropy,laflorencie16entropy,vidmar2017entropy,nahum17entropy}. Note that the lattice basis where such partition is performed is not necessarily the one in which $\mathbf{h}$ is diagonalized. For the ground states of Hamiltonians of the form \eqref{eqn:H_free}, according to Wick's theorem the expectation value of any observable is a function of the two-point correlator $\langle \Psi_N| f^\dagger_i f^{\phantom{\dagger}}_j|\Psi_N \rangle \equiv \gamma_{ji}$. The latter defines the one-particle reduced density matrix (1RDM) $\gamma$ of the ground state. As $\rho_A$ is uniquely defined via the expectations of all physical observables on the subsystem $A$\cite{friis2016reasonable}, one can then in principle express $\rho_A$ using $\gamma$.
While the explicit expression can be cumbersome, spectra of $
\rho_A$ and $\gamma$ can be easily related via the remarkable shortcut put forward by Peschel\cite{peschel2003reduced}. The shortcut makes use of two insights: (1) The reduced state $\rho_A$ is a Gaussian state $\rho_A = K \exp(\sum_{ij}C_{ij}f^\dagger_i f^{\phantom{\dagger}}_j)$, where $K$ is the normalization factor. (2) The defining matrix $\mathbf{C} = (C_{ij})$ and the 1RDM restricted to the sub-algebra on subsystem $A$, $\gamma_A$, can be simultaneously diagonalized. Combining these two ingredients, Peschel discovered that the two matrices are related via $\mathbf{C} = \ln[(\openone-\gamma_A)/\gamma_A]$.

This shortcut greatly simplifies the complexity of computing the spectrum of $\rho_A$: one simply needs to find the restricted 1RDM and its eigenvalues. As the von Neumann entropy is only a function of the spectrum, $S(\rho_A)$ can also be readily calculated, without explicit construction of $\mathbf{C}$ or $\rho_A$.
In general, however, not all correlation quantities can be determined with the spectrum of the reduced states alone. In particular, suppose $A$ contains two orthonormal orbitals. The entanglement between the two orbitals is not invariant under rotations between them and cannot be retrieved using the same method for $S(\rho_A)$. In the next Section, we will break the current paradigm by calculating explicitly the orbital-orbital entanglement in the tight-binding model, using the analytic formula\eqref{eqn:e_nssr}.

\subsection{Formula for orbital-orbital entanglement in a tight-binding model}


We consider the following tight-binding Hamiltonian for electrons on a periodic chain with $L$ lattice sites
\begin{equation}
    \hat{H}_\mathrm{tb} = -\frac{1}{2}\sum_{\sigma = \uparrow, \downarrow}\sum_{l=1}^{L} f^\dagger_{l\sigma} f^{\phantom{\dagger}}_{l+1,\sigma} + h.c. \label{eqn:H_tb}
\end{equation}
with periodic boundary condition imposed. Here, $f^{(\dagger)}_{l\sigma}$ annihilates(creates) an electron with spin $\sigma$ at the Wannier orbital/lattice site labeled by $l$. Each Wannier orbital can host up to two electrons with different spin. The discrete translation invariance of the system allows us to diagonalise the Hamiltonian via a Fourier transform
\begin{equation}
    c^\dagger_{k\sigma} = \frac{1}{\sqrt{L}} \sum_{l=1}^L e^{-\frac{2\pi i}{L}kl} f^\dagger_{l\sigma}. \label{eqn:fourier}
\end{equation}
In this basis, the Hamiltonian \eqref{eqn:H_tb} is diagonal, and the one-electron spectrum is given by the dispersion relation
\begin{equation}
    E_{k} = -\cos\left(\frac{2\pi k}{L}\right), \quad k = 0, \pm1, \ldots,\pm \Bigl\lceil\frac{L-1}{2}\Bigr\rceil,
\end{equation}
where in case of $L$ even only one of the signs $\pm \bigl\lceil\frac{L-1}{2}\bigr\rceil$ should be chosen.
As the system is non-interacting, the $N$-electron ground state is the configuration where the $N$ lowest energy levels are filled. For simplicity, we shall assume from now on that the number of electrons is either $N=0$, or $N = 4 k_\mathrm{max} + 2$ where $k_\mathrm{max} \geq 0$ denotes the highest occupied momentum. In the former case, the ground state is simply the vacuum. In the latter, the $N$-particle ground state $|\Psi_N\rangle$ is uniquely characterized by the Fermi level
\begin{equation}
    \langle \Psi_N | c^\dagger_{k\sigma} c^{\phantom{\dagger}}_{k'\sigma'} |\Psi_N\rangle = \delta_{kk'}\delta_{\sigma\sigma'}\Theta(k_\mathrm{max}-|k|), \label{eqn:fermi_sea}
\end{equation}
where $\Theta$ is the Heaviside step function. With \eqref{eqn:fourier} and \eqref{eqn:fermi_sea} at hand we are ready to compute the matrix elements of the 1RDM in the spatial basis as
\begin{equation}
    \begin{split}
        \langle \Psi_N|f^\dagger_{l\sigma} f^{\phantom{\dagger}}_{l' \sigma'} |\Psi_N\rangle = \frac{1}{L}\delta_{\sigma, \sigma'} \sum_{k=-k_\mathrm{max}}^{k_\mathrm{max}} e^{ \frac{2\pi i(l-l')k}{L} }. \label{eqn:tb_1rdm_diag}
    \end{split}
\end{equation}
When $l=l'$, all summands in \eqref{eqn:tb_1rdm_diag} become 1, and thus we obtain
\begin{equation}
    \begin{split}
        \langle \Psi_N|f^\dagger_{l\sigma} f^{\phantom{\dagger}}_{l \sigma'} |\Psi_N\rangle &= \delta_{\sigma\sigma'}\frac{2k_\mathrm{max}+1}{L}
        \\
        &= \delta_{\sigma\sigma'}\frac{N}{2L} \equiv \delta_{\sigma\sigma'}\eta,
    \end{split}
\end{equation}
where $\eta = N/(2L)$ is the particle filling fraction. This agrees with the physical intuition that the electrons move freely along the chain and appear at every site with uniform probability. When $l\neq l'$, by introducing $\omega \equiv \frac{2\pi(l-l')}{L}$, we arrive at
\begin{equation}
    \begin{split}
        \langle \Psi_N|f^\dagger_{l\sigma} f^{\phantom{\dagger}}_{l' \sigma'} |\Psi_N\rangle &= \frac{1}{L} \delta_{\sigma\sigma'}\frac{e^{i\omega k_\mathrm{max}}-e^{-i\omega k_\mathrm{max}}}{1-e^{iw}}
        \\
        &= \frac{1}{L}\delta_{\sigma\sigma'} \frac{\sin\left[\omega \left(k_\mathrm{max}+\frac{1}{2}\right)\right]}{\sin\left(\frac{\omega}{2}\right)}
        \\
        &=\frac{1}{L}\delta_{\sigma\sigma'} \frac{\sin\left(\frac{\omega N}{4} \right)}{\sin\left(\frac{\omega}{2}\right)}.
    \end{split}
\end{equation}
In the thermodynamic limit where $N,L\!\rightarrow\!\infty$ while $\eta\!=\!N/(2L)$ is kept fixed, the off-diagonal elements of the 1RDM become
\begin{equation}
\begin{split}
     \lim_{L\rightarrow\infty}\langle \Psi_N|f^\dagger_{l\sigma} f^{\phantom{\dagger}}_{l' \sigma} |\Psi_N\rangle  = \frac{\sin( \pi d \eta)}{\pi d} \equiv W(d,\eta), \label{eqn:w}
\end{split}
\end{equation}
where $d\!=\!|l-l'|$ is the orbital separation distance.

\begin{figure}[t]
    \centering
    \includegraphics[scale=0.30]{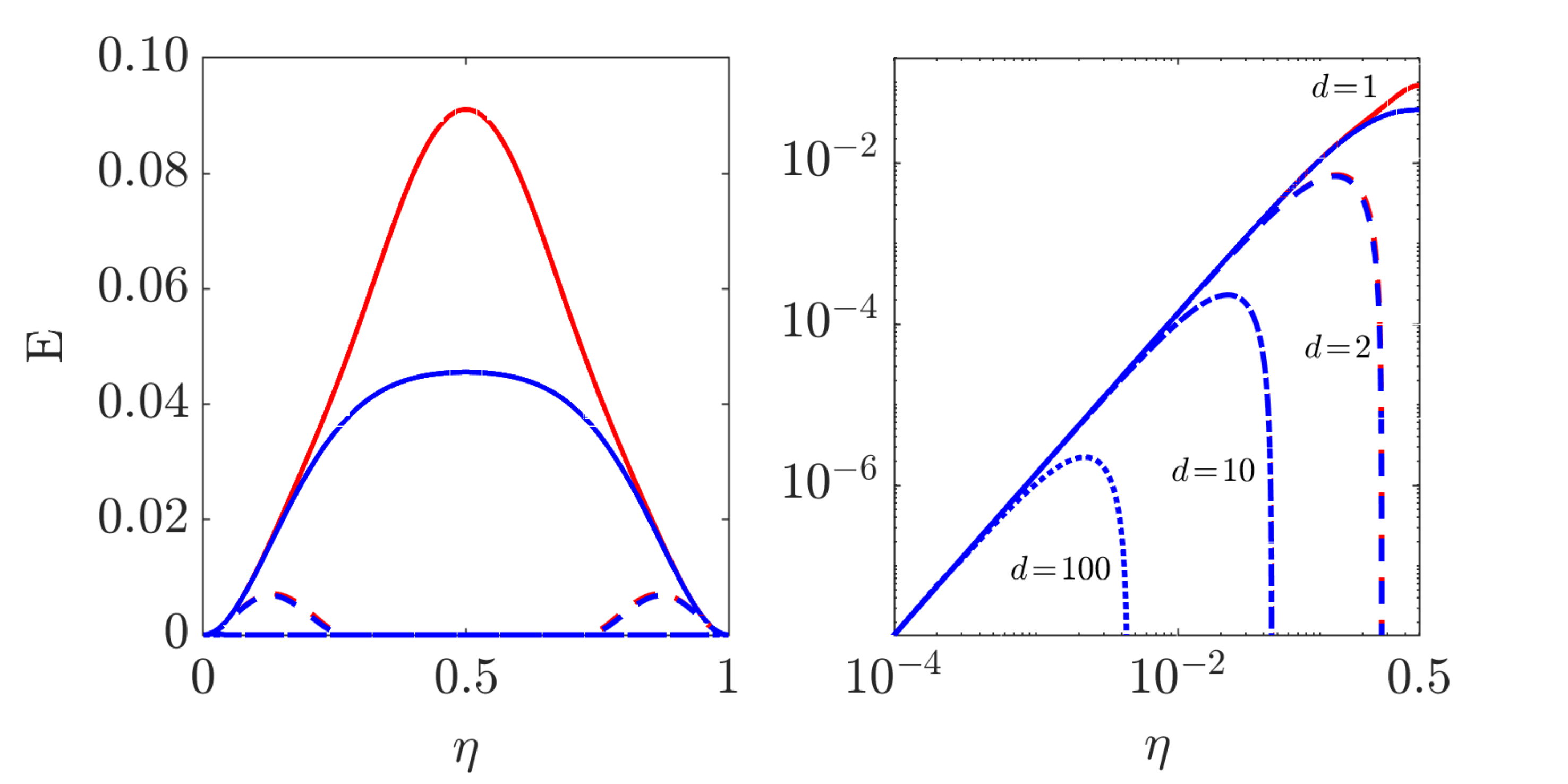}
    \caption{Entanglement between Wannier orbitals/lattice sites separated by inter-orbital distance 1 (solid), 2 (dashed), 10 (dashed-dotted), and 100 (dotted), under P-SSR (red) and N-SSR (blue) in the ground state of the tight-binding Hamiltonian \eqref{eqn:H_tb} as a function of the particle/hole density $\eta$. In the left panel, the dashed-dotted and dotted curves look essentially flat. They can be seen more clearly in the $\log$-$\log$ plot in the right panel.}
    \label{fig:free_EvsEta_combined}
\end{figure}

Next, we use Wick's theorem to calculate the quantities needed for the entanglement between the orbitals $l$ and $l'$ from the elements of the 1RDM derived above. The relevant operators for calculating $r$ and $t$ in the analytic formula \eqref{eqn:e_nssr} are the orthogonal projection $\hat{P}_{--}^{\,ll'}$ onto the local odd parity subspace, and the orthogonal projections $\hat{P}^{\,ll'}_{\Psi_{\pm}}$ onto $|\Psi_\pm\rangle$, all defined on the orbitals $l$ and $l'$
\begin{equation}
    \begin{split}
        \hat{P}_{--}^{\,ll'} &= \sum_{i,j=l,l'}\sum_{\sigma\sigma'} \hat{n}_{i\sigma}\hat{n}_{j\sigma'}(1-\hat{n}_{i\bar{\sigma}})(1-\hat{n}_{j\bar{\sigma}'})
        \\
        \hat{P}^{\,ll'}_{\Psi_{\pm}}&=|\Psi^{ll'}_{\pm}\rangle\langle\Psi^{ll'}_{\pm}|
        \\
        &= \hat{n}_{l \uparrow} \hat{n}_{l' \downarrow}(1 \!-\! \hat{n}_{l \downarrow}) (1\!-\!\hat{n}_{l' \uparrow}) \pm f^{\dagger}_{l\uparrow} f^{\dagger}_{l'\downarrow} f^{\phantom{\dagger}}_{l'\uparrow} f^{\phantom{\dagger}}_{l\downarrow}.
    \end{split}
\end{equation}
Here, $\hat{n}_{l\sigma}\! =\! f^\dagger_{l\sigma} f^{\phantom{\dagger}}_{l\sigma}$ denotes the spin-$\sigma$ particle number operator on orbital $l$. After some straightforward but lengthy calculation, we obtain the following expressions for the parameters $r$ and $t$, required in Eq.~\eqref{eqn:e_nssr},
\begin{equation}
    \begin{split}
        t &= \max\{ \langle \Psi_N |\hat{P}^{\,ll'}_{\Psi_{\pm}} |\Psi_N\rangle \} = A + B,
        \\
        r &= \langle \Psi_N | \hat{P}_{--}^{\,ll'} |\Psi_N\rangle - t = 3A - 3B,
    \end{split}
\end{equation}
where
\begin{equation}
    A \equiv \left[\eta^2 - \eta - W(d,\eta)^2 \right]^2, \quad B \equiv W(d,\eta)^2. \label{eqn:AB}
\end{equation}
With this, the N-SSR entanglement can be neatly expressed as
\begin{equation}
\begin{split}
    E(\tilde{\rho}^\mathrm{N}_{ll'}) &= (A+B)\log\left(\frac{A+B}{2A-B}\right)
    \\
    & \quad+ (3A-3B)\log\left(\frac{3A-3B}{2A-B}\right) \label{eqn:tb_ent}
\end{split}
\end{equation}
provided the entanglement criteria \eqref{eqn:ent_criteria} (here it translates to $A < 2B$) is met, and otherwise it is zero. We present in Figure \ref{fig:free_EvsEta_combined} the orbital-orbital entanglement as a function of the filling fraction $\eta$, at various orbital separations $d$, together with the entanglement when only P-SSR is assumed, also obtained analytically (formula can be found in Ref.~\cite{ding2022quantifying}).

\subsection{Discussion of orbital-orbital entanglement in tight-binding model}

In this section we discuss the orbital-orbital entanglement in the tight-binding model based on \eqref{eqn:tb_ent}. Figure \ref{fig:free_EvsEta_combined} reveals the rich entanglement structure of the Slater determinant ground state. At different particle filling fraction $\eta$, the orbital-orbital entanglement behaves in qualitatively distinct ways, which we will now account for in detail.

First, we notice that the graph of orbital-orbital entanglement is symmetric around the half-filling point $\eta\!=\!1/2$. This is due to the unbroken particle-hole symmetry of the Hamiltonian \eqref{eqn:H_tb} in the ground states. Moreover, using \eqref{eqn:tb_ent}, it is easy to prove that the entanglement as a function of $\eta$ is symmetric about $\eta\!=\!1/2$. Because of this, it suffices to consider only $0 \! \leq \! \eta \! \leq \! 1/2$, and treat it as a particle or hole filling fraction. In order to illustrate more details of this $\eta$-regime, we use a $\log$-$\log$ scale in the right panel of Figure \ref{fig:free_EvsEta_combined}.

Second, the analytic properties of the orbital-orbital entanglement highly depend on the combination of orbital separation $d$ and the filling fraction $\eta$. The common feature here is the vanishing of orbital-orbital entanglement for all $d$ at both, $\eta\!=\!0$ and $\eta\!=\!1$. In these two cases the ground state is invariant under any orbital rotation and contains zero correlation in any orbital basis.
For two neighboring orbitals ($d\!=\!1$), their entanglement is maximized exactly when the chain is half filled $(\eta\!=\!1/2)$, for both P- and N-SSR entanglement. Around half-filling, the P-SSR entanglement is significantly higher than the N-SSR one. This is not surprising, as N-SSR imposes a stronger restriction on the accessible entanglement. Yet, at small particle or hole filling fractions, the two become essentially the same. The reason for this is that in such regimes the weight in the $P_{1,1}$ sector dominates over the weight in the $P_{2,0} + P_{0,2}$ sector, as the particles/holes tend to avoid sitting on the same orbital due to the repulsive interaction. This weight domination effect is stronger when the orbitals are further apart.
For $d\!\geq\!2$, the P- and N-SSR entanglement is practically indistinguishable. Thus we can restrict the investigation of analytic properties to the latter by referring to \eqref{eqn:tb_ent} for all practical purposes here, and also in the following section. In this case, two maxima occur which appear symmetrically on the $\eta$-axis about $\eta\!=\!1/2$ (see the left panel of Figure \ref{fig:free_EvsEta_combined})  due to the particle-hole symmetry of \eqref{eqn:tb_ent}. At lower $\eta$, more and farther separated orbital pairs become entangled, though this entanglement grows vanishingly small with the filling fraction.

Third, we find that entanglement across extremely long distances can be engineered by tuning the filling fraction to be infinitesimally small. In the right panel of Figure \ref{fig:free_EvsEta_combined}, we see more clearly on the $\log$-$\log$ scale that for each separation $d$, there exists a critical value $\eta(d)$, below which the two orbitals are always entangled. At the same time, for any filling fraction $\eta$, there is a minimal disentangling distance $d_\mathrm{\min}(\eta)$, beyond which all orbital pairs are in a separable state. This phenomenon of completely vanishing entanglement is called the sudden death of entanglement\cite{yu2009entanglement}. Geometrically, it means nothing else than that the two-orbital reduced quantum state evolves along a trajectory which at some point enters the convex set of separable states. We shall explore this phenomenon in more details in the next section.

\begin{figure}[t]
    \centering
    \includegraphics[scale=0.30]{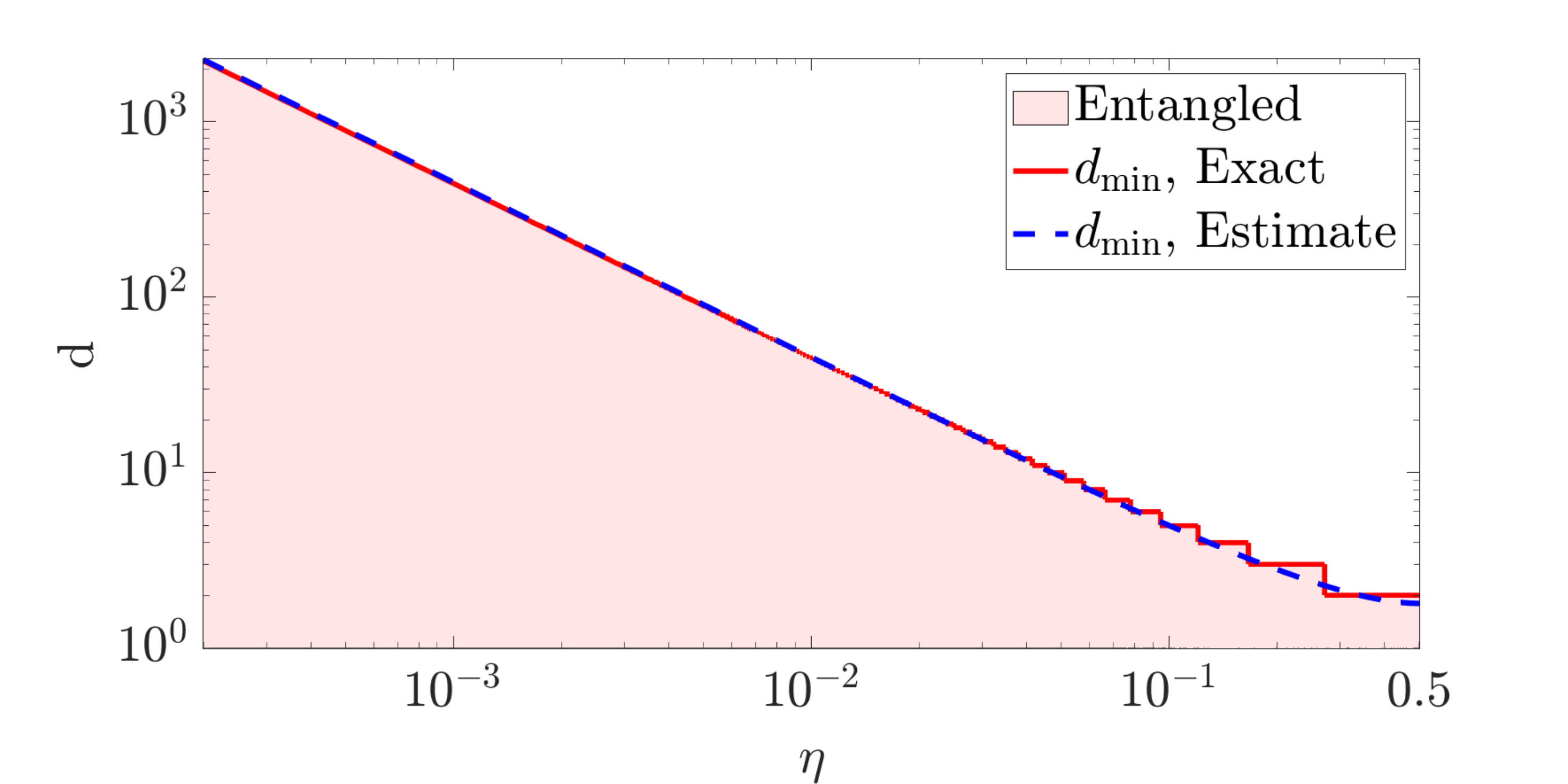}
    \caption{Disentangling distance $d_\text{min}$ as a function of particle/hole density $\eta$.}
    \label{fig:free_DminVsEta}
\end{figure}

\subsection{Limiting regimes and sudden death of entanglement}

In this section we investigate the asymptotic behavior of the orbital-orbital entanglement in various limiting regimes. Previously we observed that interesting phenomena such as long-distance entanglement and sudden death of entanglement occur when the particle filling $\eta$ or hole filling $1\!-\!\eta$ is small. Moreover, the P- and N-SSR entanglement in this region are practically identical. Therefore we can focus on the latter in our analysis.

For a fixed orbital separation $d$, when $\eta \ll d^{-1}$, we get from \eqref{eqn:w} $W(d,\eta) = \eta + \mathcal{O}(\eta^3)$. The N-SSR entanglement between two orbitals separated by $d$ becomes
\begin{equation}
    \begin{split}
        E(\tilde{\rho}^\mathrm{N}_{ij}) = 2\log(2) \eta^2 + \mathcal{O}(\eta^5), \quad \eta \ll d^{-1}, \label{eqn:ent_eta_scaling}
    \end{split}
\end{equation}
i.e., the N-SSR orbital-orbital entanglement is quadratic in $\eta$ for small $\eta$. In log-log scale this relation becomes linear
\begin{equation}
    \log(E(\tilde{\rho}^\mathrm{N}_{ij})) = 2\log(\eta) + \mathcal{O}(1), \quad \eta \ll d^{-1}. \label{eqn:tb_ent_n_asym}
\end{equation}
For the sake of completeness, we also verify that the P-SSR entanglement indeed reduces to the N-SSR one, i.e.
\begin{equation}
    \log(E(\tilde{\rho}^\mathrm{P}_{ij})) = 2 \log(\eta) + \mathcal{O}(1), \quad \eta \ll d^{-1}. \label{eqn:tb_ent_p_asym}
\end{equation}
Eqs.~\eqref{eqn:tb_ent_n_asym} and \eqref{eqn:tb_ent_p_asym} explain the asymptotic linear behavior in Figure \ref{fig:free_EvsEta_combined} (right panel) of both the P-SSR (red) and N-SSR (blue) entanglement on the $\log$-$\log$ scale, with slope 2 valid for any separation $d$. This relation is more robust for small separation $d$, as for large separation the orbital-orbital entanglement quickly deviates from the linear asymptote and plunges into sudden death as $\eta$ increases. On the other hand, for small $\eta$ and $d < d_\mathrm{min}(\eta)$, the orbital-orbital entanglement is only a function of $\eta$, but not of $d$. All orbital pairs below the critical separation are equally entangled with one another. Due to the particle-hole symmetry, the orbital-orbital entanglement at small hole fraction $1\!-\!\eta \ll d^{-1}$ is also quadratic in $1\!-\!\eta$.

Above the critical separation $d_\mathrm{min}(\eta)$, all orbital pairs are disentangled. Based on Figure \ref{fig:free_DminVsEta}, $d_\mathrm{min}(\eta)$ seems to diverge as $\eta$ approaches 0. To understand the asymptotic behavior of $d_\mathrm{min}(\eta)$ at small $\eta$, we analyze the entanglement criteria \eqref{eqn:ent_criteria}. Recall from the previous section that the two orbitals are in a separable state if $A$ and $B$ in \eqref{eqn:AB} satisfy the relation $A \geq 2B$, which in explicit terms reads
\begin{equation}
    \begin{split}
    \eta^2-\eta &\leq W(d,\eta)^2 - \sqrt{2} |W(d,\eta)|. \label{eqn:sep_criterion}
    \end{split}
\end{equation}
Although \eqref{eqn:sep_criterion} cannot be solved analytically, in the limit of large $d$ using the expansion $W(d,\eta) \sim (\pi d)^{-1}$ we can derive an analytic estimation for the minimal disentangling separation $d_\mathrm{min}(\eta)$
\begin{equation}
    d_{\min}(\eta) = \frac{\sqrt{2}}{\pi} \frac{1}{\eta(1\!-\!\eta)} + \mathcal{O}(\eta^0). \label{eqn:dmin}
\end{equation}
In $\log$-$\log$ scale, this relation becomes linear
\begin{equation}
\begin{split}
    \log(d_\mathrm{min}(\eta)) = \begin{cases}
    -\frac{\sqrt{2}}{\pi}\log(\eta) + \mathcal{O}(\eta),&\quad\:\:\eta \ll 1,
    \\
    -\frac{\sqrt{2}}{\pi}\log(1\!-\!\eta) + \mathcal{O}(1\!-\!\eta),&\,1\!-\!\eta \ll 1.
    \end{cases}
\end{split}
\end{equation}
In Figure \ref{fig:free_DminVsEta}, we plot the exact minimal disentangling distance obtained by numerically locating the sudden death of entanglement using \eqref{eqn:sep_criterion}, together with the analytic estimation \eqref{eqn:dmin}. We can see the agreement is excellent, even up to the order $10^{-1}$ in $\eta$.

To summarize Section \ref{sec:freefermions}, building on Ref.~\cite{ding2022quantifying}, we derived an analytic expression for the entanglement between any two spatial orbitals in the tight-binding model \eqref{eqn:H_tb}, as a function of both inter-orbital separation $d$ and particle filling fraction $\eta$. In particular, Eq.~\eqref{eqn:e_nssr} revealed the existence of long-distance entanglement when the filling fraction is low or close to 1. Moreover, our asymptotic analysis for small $\eta$ revealed that this entanglement between orbitals is almost independent of the orbital separation $d$, and quadratic in the particle filling $\eta$ for small $\eta$ (and quadratic in hole filling $1-\eta$ for small $1-\eta$). For a fixed filling $\eta$, the entangling range is still finite, as we also observed the sudden death of entanglement when the two orbitals are separated beyond a critical distance. Utilizing our analytic expression, the leading order of the minimal disentangling distance $d_\mathrm{min}$ is extracted to be diverging as $\eta^{-1}$ when $\eta$ is small, and $(1\!-\!\eta)^{-1}$ when $\eta$ is close to 1.

\begin{figure}[t]
    \centering
    \includegraphics[scale=0.40]{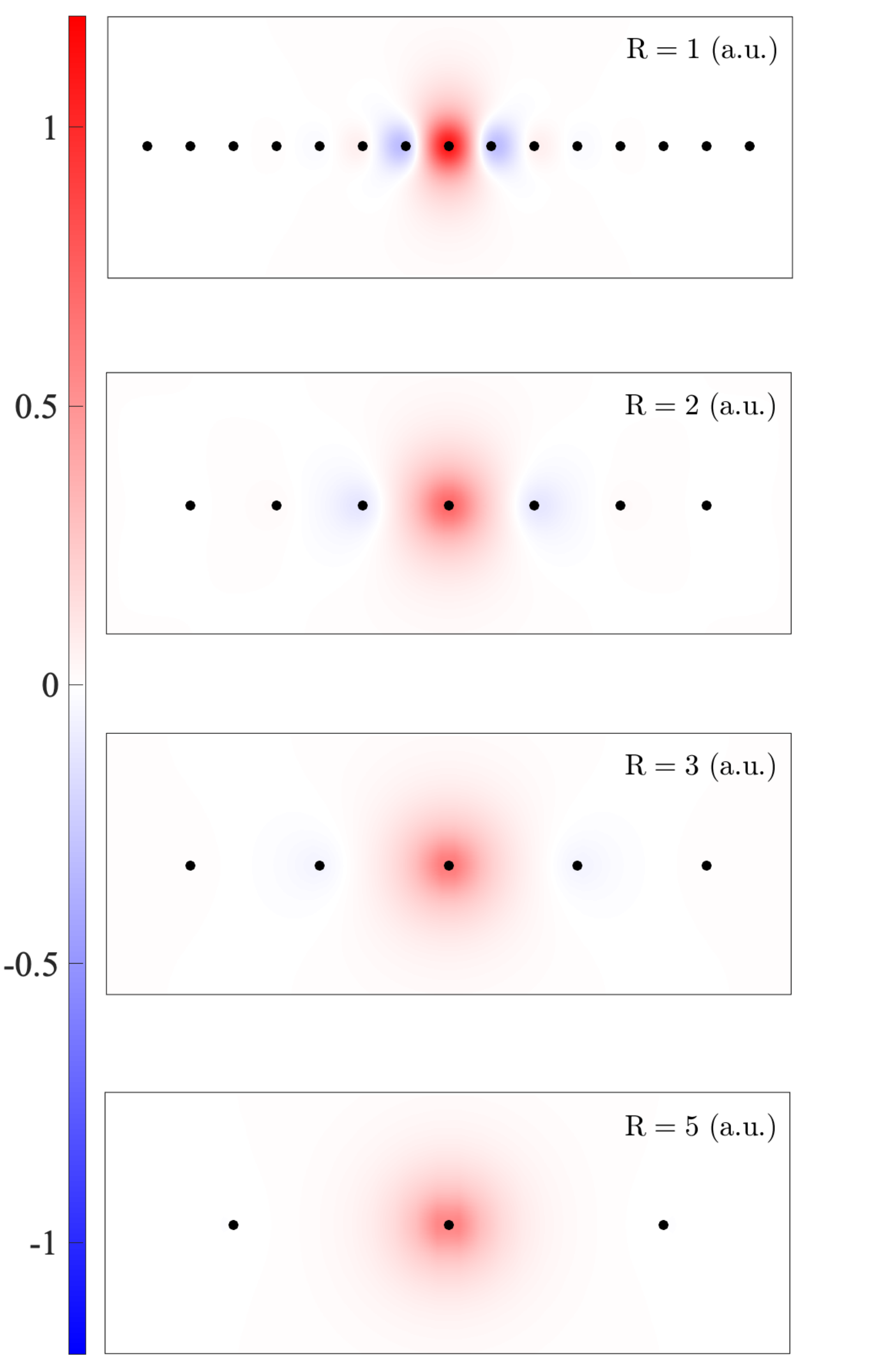}
    \caption{Values of the localized orbitals' wavefunctions in $\mathrm{H_{16}}$ in the STO-3G basis, for various nearest neighbor distances $R$ (a.u.). Nuclear centers are represented as black dots. The horizontal range is 16 a.u., and the vertical range 6 a.u..}
    \label{fig:h16_orbitals}
\end{figure}

\section{Hydrogen ring}

\begin{figure*}[t]
    \centering
    \includegraphics[scale=0.33]{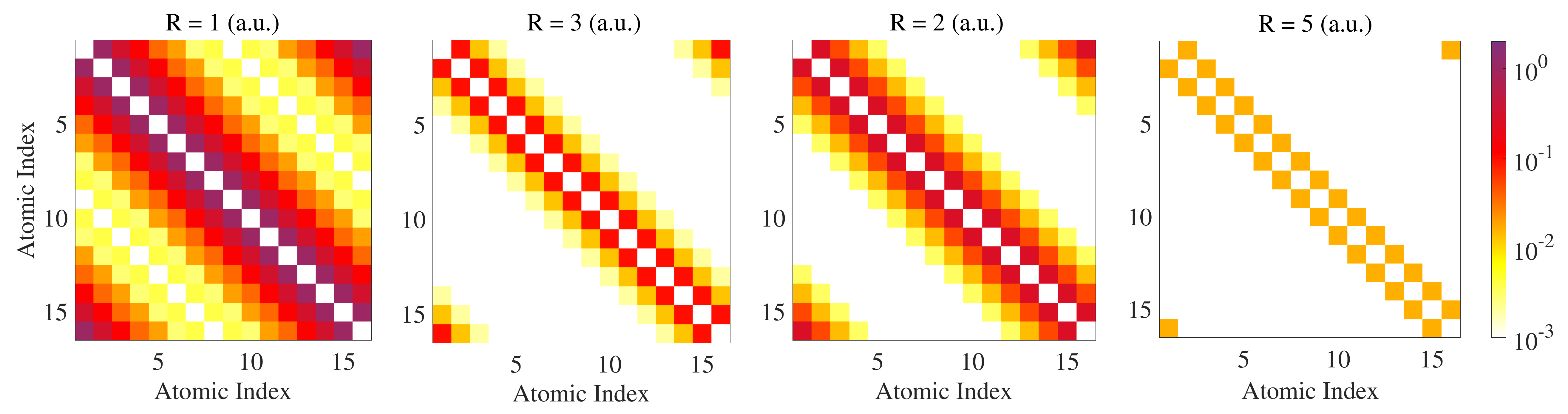}
    \caption{Values of matrix elements of the one-electron Hamiltonian $|(h_1)_{ij}|$ (a.u.) with $i\!\neq\!j$, for nearest neighbor separation $R=1,2,3,5$ (a.u.). Periodic boundary condition is imposed by the ring geometry.}
    \label{fig:h16_1ptH}
\end{figure*}

\label{sec:hydrogen}

In the previous section, we thoroughly explored the orbital-orbital entanglement of the tight-binding model \eqref{eqn:H_tb} for free electrons, using the analytic tools provided in Ref.~\cite{ding2022quantifying}. This rises the question of how interactions between electrons influence the orbital-orbital entanglement. Therefore we study in this section the hydrogen ring $\mathrm{H}_{16}$. In order to motivate more the choice of that system, we recall that the hydrogen ring (or open chain) is a common system for benchmarking ground state methods\cite{hchain1992,wright1992structure,hchain2011,hchain2017,hchain2023}. This is due to the fact that it contains a significant amount of entanglement and thus aims at simulating electron correlation effects in realistic materials. Yet, the strong correlation therein presents a significant numerical challenge. Moreover, this and related ring models can be experimentally realized in the context of ultracold atom experiments\cite{Amico2005,Franke2007,Ramanathan2011,Amico2014,Bell2016}.

The hydrogen ring shares many similarities with the tight-binding chain: a localized orbital basis can be established, and there is a local hopping generated by the overlapping atomic orbitals. As a molecular system, however, it additionally contains a periodic nuclear potential, and the Coulomb interaction between the electrons which is the source of computational complexity. It is thus of considerable interest to ask to what extend the entanglement effects in the hydrogen ring can be explained by the insights we gained from the tight-binding model \eqref{eqn:H_tb} of non-interacting electrons in the previous section. Although the orbital-orbital correlation structure of similar systems has been analyzed\cite{legeza2014}, the orbital-orbital \textit{entanglement} structure is yet to be determined.

\begin{figure*}[t]
    \centering
    \includegraphics[scale=0.49]{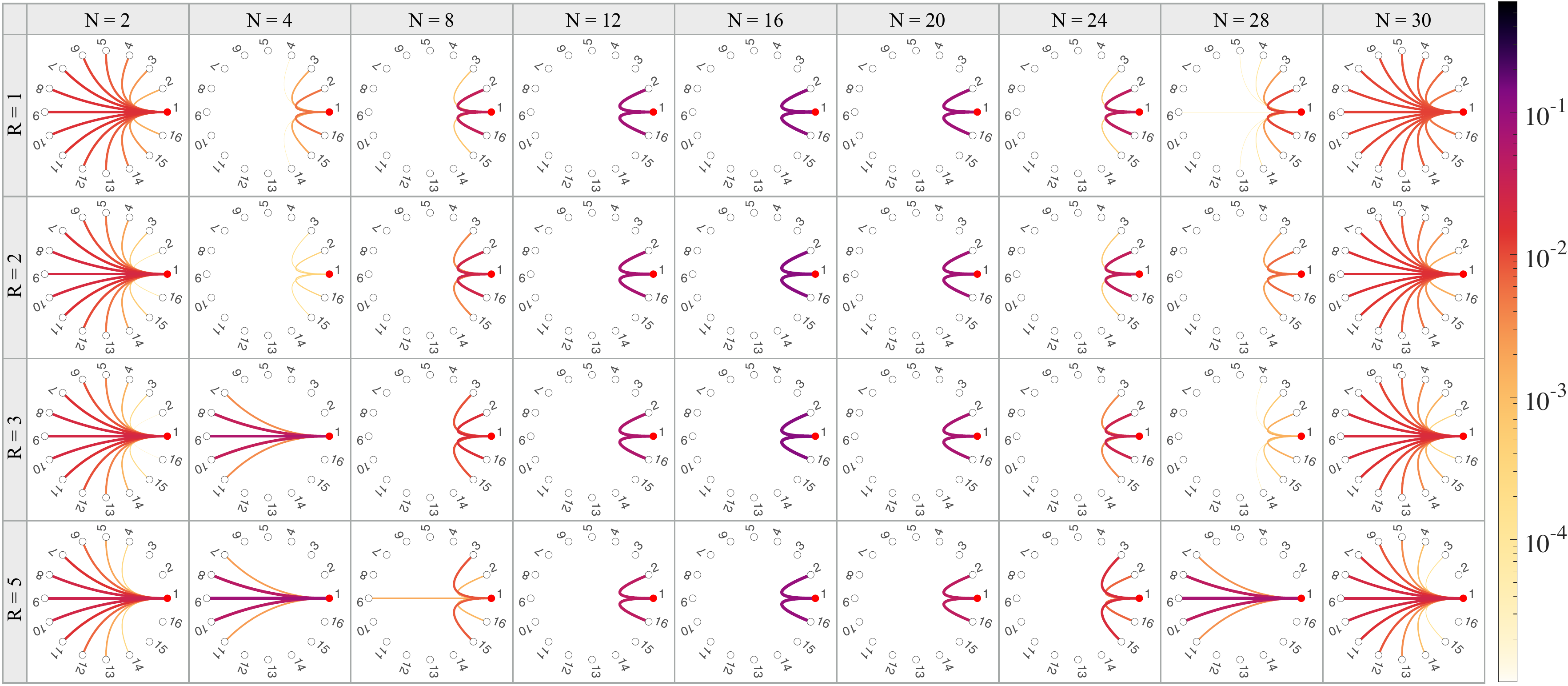}
    \caption{P-SSR entanglement between localized orbitals in $\mathrm{H_{16}}$ in STO-3G basis, for various numbers $N$ of electrons and nearest neighbor distances $R$ (a.u.). Since all orbitals are identical up to relabeling, only entangled pairs involving orbital 1 (filled red circle) are shown.}
    \label{fig:h16_ent_pssr}
\end{figure*}

\subsection{Model and Hamiltonian}

We consider a finite periodic hydrogen ring, defined by the uniform nearest internuclear distance $R$, with the minimal STO-3G basis set. This basis contains one 1s orbital at each atomic center and it is the minimal setting for describing the dissociation of hydrogen molecules.

The electronic Hamiltonian contains three terms: the nuclear potential, the kinetic energy, and the Coulomb interaction. The corresponding Hamiltonian contains one- and two-electron terms $\hat{h}$ and $\hat{V}$, respectively, and takes the form
\begin{equation}
\begin{split}
    &\hat{H} = \hat{h} + \hat{V}
    \\
    &\:\:\:= \sum_{ij}\sum_\sigma {h}_{ij} f^\dagger_{i\sigma} f^{\phantom{\dagger}}_{j\sigma} + \sum_{ijkl} \sum_{\sigma\sigma'} V_{ijkl} f^\dagger_{i\sigma} f^\dagger_{j\sigma'} f^{\phantom{\dagger}}_{l\sigma'}f^{\phantom{\dagger}}_{k\sigma},
\end{split}
\end{equation}
where
\begin{equation}
    \begin{split}
        h_{ij} &= \int \mathrm{d}x^3 \phi_{i}^\ast(x) \left( -\frac{\hbar}{2m_e} \nabla^2 + \sum_{m=0}^{M-1} \frac{Z_m}{|x-\mathbf{X}_m|} \right) \phi_j(x),
        \\
        V_{ijkl} &=  \frac{1}{2}\int  \mathrm{d}x^3  \mathrm{d}x'^3  \phi_{i}^\ast(x) \phi_{j}^\ast(x') \frac{1}{|x-x'|} \phi_{l}(x') \phi_k(x).
    \end{split}
\end{equation}
Here, $\phi_i$'s are orthonormal molecular orbitals which form a complete basis of the one-particle Hilbert space spanned by the non-orthogonal atomic orbitals. $M$ is the total number of nuclear centers with charges $Z_m$. In the case of $\mathrm{H}_{16}$, $Z_m\!=\!1$ for all $m\!=\!1,2,\ldots,16$ and $\mathbf{X}_m$ are defined as
\begin{equation}
\begin{split}
    \mathbf{X}_{m} = \frac{R}{2\sin\left(\frac{\theta}{2}\right)} (\cos(m\theta),\sin(m \theta),0), \quad \theta= \frac{2\pi}{16},
\end{split}
\end{equation}
so that the 16 hydrogen atoms are arranged periodically on a ring, with the straight-line distance between two nearest neighbor atoms equal to $R$. From here on, we shall use atomic units (denoted by a.u.) for both distance and energy, which correspond to Bohr and Hartree, respectively.

The matrix elements of the one- and two-electron Hamiltonians are explicitly orbital basis dependent. For the purpose of QIP, we choose a set of localized orthonormal orbitals, obtained by symmetrically orthogonalizing\cite{lowden70} the local 1s atomic orbitals. In Figure \ref{fig:h16_orbitals}, we plot these localized orbitals for different nearest neighbor distances $R$. Locally they resemble the atomic orbitals, but they necessarily have finite contributions from atomic orbitals on other nuclear centers to ensure orthogonality. At $R\!=\!1$, each localized orbital at the corresponding atomic center still contains significant weight of the atomic orbitals on the nearest neighbors, while for $R \!\geq\! 3$, the localized orbitals are virtually single-centered.

Do these localized orbitals exhibit a local interaction structure similar to the tight-binding chain? To answer this question, we plot the hopping terms $h_{ij}$ for $i\!\neq\! j$ in Figure \ref{fig:h16_1ptH}. Recall that the hydrogen atoms are arranged on a ring, and therefore orbital 1 and 16 are in fact nearest neighbors, despite appearing on the opposite ends of the axes in Figure \ref{fig:h16_1ptH}. At small separation, e.g. $R\!=\!1$ (a.u.), hopping between distant localized orbitals is possible, but the hopping strength decays exponentially as internuclear distance increases. As we stretch the hydrogen ring, the hopping strength decays faster with the orbital separation. Finally at $R\!=\!5$ (a.u.), the hopping is strictly between nearest neighbors, just as in the tight-binding model \eqref{eqn:H_tb}.

\subsection{Numerical methods}

The calculation of the ground state of the hydrogen ring is done in three steps: (1) using the atomic wavefunction overlap matrix provided by \textsc{Molpro}\cite{molpro1,molpro2,molpro3}, we symmetrically orthogonalize\cite{lowden70} the atomic orbitals into the localized orbitals, for nearest neighbor separation $R \!=\! 1,2,3,5$ (a.u.). (2) With a preceding Hartree-Fock calculation, we then transform the Hartree-Fock canonical orbitals into the localized orbitals as a post Hartree-Fock step, and compute the matrix elements of the one- and two-electron Hamiltonians followed by outputting them in \texttt{FCIDUMP}\cite{know89b} file format, all within the framework of \textsc{Molpro}. (3) With the \texttt{FCIDUMP} file, quantum chemistry density matrix renormalization group (QC-DMRG) calculations are performed to solve for the ground states, using the \textsc{SyTen}\cite{hubig:_syten_toolk,hubig17:_symmet_protec_tensor_networ} package, originally created by Claudius Hubig. The calculations are carried out for various electron numbers, initialized with random configuration states. The DMRG convergence is particularly challenging, because of both the strong intrinsic correlation in the system and the additional entanglement due to the localized orbital basis representation. For the data we use, the maximal bond dimension is set at $m\!=\!1000$. The final energies are chemically accurate, verified by the energy improvement below the order $10^{-3}$ (a.u.) in the DMRG calculations performed with bond dimension 2000.

\subsection{Plots and results}

In this section we present our main findings on the entanglement structure of the hydrogen ring $\mathrm{H}_{16}$. We tabulate in Figure \ref{fig:h16_ent_pssr} the P-SSR entanglement between two localized orbitals for various nearest neighbor separations $R$ and different numbers of electrons $N$ in the ground state. For readability we exploit the discrete rotational symmetry of the system and only plot the entanglement associated to the localized orbital labeled 1 (represented as filled red circle). Entanglement involving other localized orbitals can be understood by relabeling them. Exact values of the P- and N-SSR entanglement can be found in Table \ref{tab:h16} in Appendix \ref{app:tables}. With these data, we observe three consequential features of the orbital-orbital entanglement between the localized orbitals in $\mathrm{H}_{16}$.

First, we observe that at half-filling ($N\!=\!16$), the orbital-orbital entanglement is in complete agreement with the tight-binding model \eqref{eqn:H_tb}, namely only neighboring localized orbitals are entangled. In fact, this agreement extends beyond half-filling, as there is a region of filling fraction around $1/2$ where all orbital pairs separated by more than one lattice constant are disentangled, exactly as in Figure \ref{fig:free_EvsEta_combined}. This is particularly surprising for systems with small nearest neighbor distance $R$, since there the hopping strength beyond nearest neighbors is rather considerable.

Second, at low or high filling, the hydrogen ring also displays long-distance entanglement similar to the tight-binding model. This phenomenon here differs from the latter, however, by preferring to entangle localized orbitals that are farther apart. Recall that in the tight-binding model, the orbital-orbital entanglement is uniform when long-distance entanglement is realized. This departure from the non-interacting scenario is best exemplified by the $N\!=\!2$ column in Figure \ref{fig:h16_ent_pssr}. In the ground state, due to the strong Coulomb repulsion, the two electrons would like to situate themselves as far apart as possible, forming singlet pairs at distant localized orbitals. Singlet states of such type, usually observed in a dissociated molecule, contain significant amount of entanglement. Singlet states realized on orbitals located closer to each other are of course possible, but they contribute to the total ground state with smaller weights. These two observations put together qualitatively explain the system's preference for long-distance entanglement at low particle filling. Furthermore, we notice that despite the lack of particle-hole symmetry in the system, the orbital entanglement in the $N\!=\!30$ column in Figure \ref{fig:h16_ent_pssr} displays a similar behavior as in $N\!=\!2$. This can be qualitatively understood with the observation that at very low (or very high) filling, the particle-particle (or hole-hole) interaction becomes negligible compared to the one-particle terms. In the extreme case where the system hosts only one particle or one hole, the corresponding Hamiltonians are related to each other by a minus sign. Consequently, the two ground states are identical up to a global phase, and therefore share the same orbital entanglement pattern. For orbitals separated by large distances, this near particle-hole symmetry is even more apparent.

Third, the long-distance entanglement in the hydrogen ring is far more potent than that in the tight-binding model. In Eq.~\eqref{eqn:ent_eta_scaling}, we see that the long-distance entanglement in the tight-binding chain is quadratic in $\eta$ for small $\eta$, which is obviously not the case here. This is best demonstrated by the entanglement between orbitals 1 and 9, which form the farthest separated pair, at $N\!=\!2$. The entanglement between them is of the order $10^{-2}$, which is the same order as the nearest neighbor entanglement. In the tight-binding model, the available entanglement at this filling fraction would yield at the order $10^{-3}$. This yet again validates that interacting electronic systems contain significant entanglement, and the entanglement is shown to be naturally distributed across distant regions, making it ideal for realizing quantum information protocols.

To summarize the present section, we took the interaction between electrons into account by studying the hydrogen ring, and analyzed the orbital-orbital entanglement therein. We discovered that the entanglement between the localized orbitals on the hydrogen ring can to a large degree be rationalized using the much simpler tight-binding model: close to half-filling only nearest neighbor orbitals can be entangled, but away from half-filling, long-distance entanglement appears. Yet, in contrast to the vanishingly weak long-distance entanglement in the tight-binding model, the entanglement between two localized orbitals sitting on far-separated nuclei is much more potent, at an order comparable to the nearest neighbor entanglement. This surprising feature, conjectured to be a result of Coulomb interaction, suggests that chain-like molecules can indeed be considered as reservoirs of entanglement.

\section{Summary and conclusions}

It was the goal of our work to guide the development of quantum technologies in the era of the ongoing second quantum revolution. For this, we identified three key aspects which are beneficial if not indispensable for the successful utilization of orbital entanglement in atoms and molecules for quantum information processing (QIP): (i) orbital localization, (ii) operationally meaningful entanglement quantification, and (iii) computable entanglement measures. To elucidate those three aspects, we first highlighted the necessity of superselection rules (SSR) with an entanglement swapping protocol. This confirmed that only the SSR-compatible entanglement in fermionic systems can be extracted from individual molecules and used for QIP tasks. We then forwent the commonly used block entropy and calculated by analytical means the physical entanglement between spatial orbitals for the exactly solvable tight-binding model. In that sense, we also overcame the restriction of previous entanglement analyses to unitarily invariant settings, i.e.~bipartitions of the system into macroscopically large and thus impractical subsystems. Finally, a numerically exact study of the hydrogen ring confirmed that the same entanglement analysis can be realized in \emph{continuous} molecular systems, provided a scheme for localizing the molecular orbitals is implemented.


In the tight-binding model, we found a rich entanglement structure despite its simplicity. Around half-filling, the entanglement is strictly local, i.e., orbitals are entangled only if they are located at neighboring atomic centers. At very low and very high particle filling, however, the system displays long-distance entanglement between orbitals, even though the hopping is strictly between nearest neighbors. To investigate this phenomenon, we extracted the asymptotic behavior of this long-distance entanglement, and found it to be quadratic in the filling fraction $\eta$ at small values of $\eta$. This low entanglement is shared uniformly within the entangling distance. The size of the entangling region, characterized by the minimal disentangling distance $d_\mathrm{min}$, diverges as the inverse of $\eta$.

For the interacting hydrogen ring $\mathrm{H}_{16}$ we solved for the ground state in the localized orbital basis. This process was by itself a numerical challenge, for both the intrinsic complexity of the interacting ground state, and the extrinsic complexity from the localized orbital representation. After obtaining the ground state, we found the entanglement pattern to be similar to the tight-binding model. Only nearest neighbors are entangled near half-filling and long-distance entanglement appears as the filling approaches 0 or 1. What distinguishes the hydrogen ring from the tight-binding model is that in the former the long-distance entanglement is much more potent. In the hydrogen ring, the ground state actually prefers to place more orbital-orbital entanglement in far separated orbital pairs, due to the Coulomb repulsion between the electrons. This effect becomes evident as the molecule starts to dissociate and the entanglement between two localized orbitals sitting at the opposite sides of the ring grows stronger, making the hydrogen ring model a more suited one for preparing and storing long-distance entanglement.

Our results confirm that molecular systems offer indeed excellent prospects for providing valuable entanglement resources required in QIP tasks. At the same time, they invite a systematic exploration by both communities of quantum science and technology, and of quantum chemistry, under the framework put forward in this work. Open challenges include tailoring an orbital localization scheme that maximizes the orbital entanglement and in that sense maximizes the resourcefulness of molecular systems\cite{ding2022quantum}. Another crucial point is to take into account the \emph{practical} restrictions on operations and measurements on localized orbitals\cite{krylov20oribitals}, resulting in more inaccessible coherent superposition terms in the two-orbital reduced state, similar to the effect of the parity or particle number SSRs. Such technical limitations enforce an effective SSR. Based on this, the actual entanglement resource available in atoms and molecules will need to be further reassessed, following the same treatment as in this work. Finally, we anticipate the challenge of extending the scope of our entanglement analysis to the case of multiple copies of the molecular system. There, the local operations do not need to respect the SSR on the individual orbitals anymore but just on the collection of all local orbitals, thus offering more entanglement resource. Yet, both quantifying the entanglement in the multi-copy quantum state, and experimentally manipulating many orbitals at the same time, are immensely difficult tasks to accomplish.

\bigskip


\begin{acknowledgements}
We thank Z.\hspace{0.5mm}Zimborás for instructive discussions at the earliest stage of this project and D.\hspace{0.5mm}Aliverti-Piuri, K.\hspace{0.5mm}Chatterjee and J.\hspace{0.5mm}Liebert for helpful comments on the manuscript. We acknowledge financial support from the Deutsche Forschungsgemeinschaft (DFG, German Research Foundation), Grant SCHI 1476/1-1 (L.D., C.S.), the Munich Center for Quantum Science and Technology (L.D., C.S.), the Gonville \& Caius College summer internship grant (G.D.), and the Max Weber Program Bavaria (G.D.). The project/research is also part of the Munich Quantum Valley, which is supported by the Bavarian state government with funds from the Hightech Agenda Bayern Plus.
\end{acknowledgements}

\bibliography{refs}

\newpage
\onecolumngrid
\appendix
\section{P- and N-SSR orbital-orbital entanglement in $\mathrm{H_{16}}$}
\label{app:tables}

\begin{table*}[h]
    \centering
    \begin{tabular}{lllll}
    \hline
      $N$   &  $R=1$ & $R=2$ & $R=3$ & $R=5$
       \rule{0pt}{2.8ex}\rule[-1.2ex]{0pt}{0pt}\\
    \hline
    \hline
$2$ $\quad\quad$ & (1, 0.00079, 0.00079) $\quad\quad$ & (1, 0.00004, 0.00004) $\quad\quad$ & (1, 0.00001, 0.00001) $\quad\quad$ & (3, 0.00015, 0.00015)  \rule{0pt}{2.8ex}\rule[-1.2ex]{0pt}{0pt}\\
    & (2, 0.00155, 0.00155) & (2, 0.00027, 0.00027) & (2, 0.00013, 0.00013) & (4, 0.00101, 0.00101)  \rule{0pt}{2.2ex}\rule[-1.2ex]{0pt}{0pt}\\
    & (3, 0.00289, 0.00289) & (3, 0.00111, 0.00111) & (3, 0.00071, 0.00071) & (5, 0.00401, 0.00401)  \rule{0pt}{2.2ex}\rule[-1.2ex]{0pt}{0pt}\\
    & (4, 0.00481, 0.00481) & (4, 0.00308, 0.00308) & (4, 0.00244, 0.00244) & (6, 0.01019, 0.01019)  \rule{0pt}{2.2ex}\rule[-1.2ex]{0pt}{0pt}\\
    & (5, 0.00708, 0.00708) & (5, 0.00642, 0.00642) & (5, 0.00589, 0.00589) & (7, 0.01751, 0.01751)  \rule{0pt}{2.2ex}\rule[-1.2ex]{0pt}{0pt}\\
    & (6, 0.00929, 0.00929) & (6, 0.01056, 0.01056) & (6, 0.01067, 0.01067) & (8, 0.02091, 0.02091)  \rule{0pt}{2.2ex}\rule[-1.2ex]{0pt}{0pt}\\
    & (7, 0.01090, 0.01090) & (7, 0.01409, 0.01409) & (7, 0.01505, 0.01505) &   \rule{0pt}{2.2ex}\rule[-1.2ex]{0pt}{0pt}\\
    & (8, 0.01149, 0.01149) & (8, 0.01549, 0.01549) & (8, 0.01685, 0.01685) &   \rule{0pt}{2.2ex}\rule[-1.2ex]{0pt}{0pt}\\ \hline
$4$   & (1, 0.00317, 0.00315)  & (1, 0.00020, 0.00020) & (6, 0.00192, 0.00192) & (6, 0.00132, 0.00132) \rule{0pt}{2.8ex}\rule[-1.2ex]{0pt}{0pt}\\
      & (2, 0.00104, 0.00102)  & (2, 0.00007, 0.00007) & (7, 0.02821, 0.02821) & (7, 0.03748, 0.03748) \rule{0pt}{2.2ex}\rule[-1.2ex]{0pt}{0pt}\\
      & (3, 0.00002, 0.00000)  &                       & (8, 0.04705, 0.04705) & (8, 0.06498, 0.06498)  \rule{0pt}{2.2ex}\rule[-1.2ex]{0pt}{0pt}\\\hline
$8$   & (1, 0.02806, 0.02361)  & (1, 0.01858, 0.01714) & (6, 0.01120, 0.01097) & (1, 0.00065, 0.00065) \rule{0pt}{2.8ex}\rule[-1.2ex]{0pt}{0pt}\\
      & (2, 0.00036, 0.00016)  & (2, 0.00217, 0.00213) & (7, 0.00561, 0.00560) & (2, 0.00554, 0.00554) \rule{0pt}{2.2ex}\rule[-1.2ex]{0pt}{0pt}\\
      &                        &                       &                       & (8, 0.00091, 0.00000)  \rule{0pt}{2.2ex}\rule[-1.2ex]{0pt}{0pt}\\    \hline
$12$  & (1, 0.06344, 0.03441)  & (1, 0.05666, 0.03542) & (1, 0.04819, 0.03722) & (1, 0.03391, 0.03344) \rule{0pt}{2.8ex}\rule[-1.2ex]{0pt}{0pt}\\\hline
$16$  & (1, 0.09116, 0.04525)  & (1, 0.10340, 0.05546) & (1, 0.10637, 0.06732) & (1, 0.08142, 0.07892) \rule{0pt}{2.8ex}\rule[-1.2ex]{0pt}{0pt}\\\hline
$20$  & (1, 0.06569, 0.03529)  & (1, 0.06303, 0.03805) & (1, 0.05458, 0.04065) & (1, 0.03310, 0.03261) \rule{0pt}{2.8ex}\rule[-1.2ex]{0pt}{0pt}\\\hline
$24$  & (1, 0.03235, 0.02639)  & (1, 0.02917, 0.02565) & (1, 0.02084, 0.01995) & (1, 0.00418, 0.00418) \rule{0pt}{2.8ex}\rule[-1.2ex]{0pt}{0pt}\\
      & (2, 0.00024, 0.00000)  & (2, 0.00030, 0.00015) & (2, 0.00200, 0.00196) & (2, 0.01301, 0.01301) \rule{0pt}{2.2ex}\rule[-1.2ex]{0pt}{0pt}\\\hline
$28$  & (1, 0.00709, 0.00696)  & (1, 0.00380, 0.00378) & (1, 0.00078, 0.00078) & (6, 0.00169, 0.00169) \rule{0pt}{2.8ex}\rule[-1.2ex]{0pt}{0pt}\\
      & (2, 0.00142, 0.00132)  & (2, 0.00113, 0.00112) & (2, 0.00030, 0.00030) & (7, 0.03275, 0.03275) \rule{0pt}{2.2ex}\rule[-1.2ex]{0pt}{0pt}\\
      & (3, 0.00005, 0.00000)  &                       & (3, 0.00002, 0.00002) & (8, 0.05492, 0.05492)  \rule{0pt}{2.2ex}\rule[-1.2ex]{0pt}{0pt}\\
      & (4, 0.00002, 0.00000)  &                       &                       &                \rule{0pt}{2.2ex}\rule[-1.2ex]{0pt}{0pt}\\  \hline
$30$& (1, 0.00306, 0.00305) & (1, 0.00082, 0.00082) & (1, 0.00018, 0.00018) & (3, 0.00046, 0.00046)  \rule{0pt}{2.8ex}\rule[-1.2ex]{0pt}{0pt}\\
    & (2, 0.00381, 0.00381) & (2, 0.00170, 0.00170) & (2, 0.00071, 0.00071) & (4, 0.00193, 0.00193)  \rule{0pt}{2.2ex}\rule[-1.2ex]{0pt}{0pt}\\
    & (3, 0.00468, 0.00468) & (3, 0.00314, 0.00314) & (3, 0.00199, 0.00199) & (5, 0.00535, 0.00535)  \rule{0pt}{2.2ex}\rule[-1.2ex]{0pt}{0pt}\\
    & (4, 0.00559, 0.00559) & (4, 0.00506, 0.00506) & (4, 0.00417, 0.00417) & (6, 0.01064, 0.01064)  \rule{0pt}{2.2ex}\rule[-1.2ex]{0pt}{0pt}\\
    & (5, 0.00642, 0.00642) & (5, 0.00719, 0.00719) & (5, 0.00708, 0.00708) & (7, 0.01584, 0.01584)  \rule{0pt}{2.2ex}\rule[-1.2ex]{0pt}{0pt}\\
    & (6, 0.00710, 0.00709) & (6, 0.00914, 0.00914) & (6, 0.01011, 0.01011) & (8, 0.01805, 0.01805)  \rule{0pt}{2.2ex}\rule[-1.2ex]{0pt}{0pt}\\
    & (7, 0.00753, 0.00753) & (7, 0.01052, 0.01052) & (7, 0.01242, 0.01242) &   \rule{0pt}{2.2ex}\rule[-1.2ex]{0pt}{0pt}\\
    & (8, 0.00768, 0.00768) & (8, 0.01102, 0.01102) & (8, 0.01328, 0.01328) &   \rule{0pt}{2.2ex}\rule[-1.2ex]{0pt}{0pt}\\ \hline
    \end{tabular}
    \caption{Tuples $(d,E_\mathrm{P},E_\mathrm{N})$ of orbital separation distance, P-SSR, and N-SSR orbital-orbital entanglement for different numbers of electrons $N$ in the ground state of $\mathrm{H_{16}}$. Only entries with $E_\mathrm{P} \geq 10^{-5}$ are shown.}
    \label{tab:h16}
\end{table*}

\end{document}